\documentclass[proceedings]{stacs}
\stacsheading{2009}{529--540}{Freiburg}
\firstpageno{529}

\usepackage{amssymb}
\usepackage{amsmath}
\usepackage{epsfig}
\usepackage{graphics}
\usepackage{color}
\usepackage[ruled]{algorithm2e}
\usepackage{pstricks}

\begin{document}

\title[Edit-Distance Computation via Text-Compression]{A Unified Algorithm for Accelerating\\ Edit-Distance Computation\\ via Text-Compression}

\author[lab1]{Danny Hermelin}{Danny Hermelin$^{*}$}\thanks{$^{*}$ Supported by the Adams Fellowship of the Israel Academy of
Sciences and Humanities.}
\author[lab1,lab4]{Gad M. Landau}{Gad M. Landau$^{\dagger}$}\thanks{$^{\dagger}$ Partially supported by the Israel Science Foundation grant 35/05
and the Israel-Korea Scientific Research Cooperation.}
\address[lab1]{Department of Computer Science,
University of Haifa, Haifa, Israel.}
\email{danny@cri.haifa.ac.il, landau@cs.haifa.ac.il}
\address[lab4]{Department of Computer and Information
Science, Polytechnic Institute of NYU, NY, USA.}
\author[lab2]{Shir Landau}{Shir Landau}
\address[lab2]{Department of Computer Science,\\
Bar-Ilan University, Ramat Gan, Israel.}
\email{shir.landau@live.biu.ac.il}
\author[lab3]{Oren Weimann}{Oren Weimann$^{\ddagger}$}\thanks{$\ddagger$ Partially supported by the Israel-Korea Scientific Research Cooperation and by the Center for Massive Data Algorithmics (MADALGO) -- a center of the Danish National Research Foundation.}
\address[lab3]{MIT Computer Science and Artificial Intelligence
Laboratory, Cambridge, MA, USA.}
\email{oweimann@mit.edu}


\begin{abstract}

The edit distance problem is a classical fundamental problem in
computer science in general, and in combinatorial pattern
matching in particular. The standard dynamic-programming
solution for this problem computes the edit-distance between a
pair of strings of total length $O(N)$ in $O(N^2)$ time. To
this date, this quadratic upper-bound has never been
substantially improved for general strings. However, there are
known techniques for breaking this bound in case the strings
are known to compress well under a particular compression
scheme. The basic idea is to first compress the strings, and
then to compute the edit distance between the compressed strings.

\hspace{10pt} As it turns out, practically all known $o(N^2)$
edit-distance algorithms work, in some sense, under the same
paradigm described above. It is therefore natural to ask
whether there is a single edit-distance algorithm that works
for strings which are compressed under any compression scheme.
A rephrasing of this question is to ask whether a single
algorithm can exploit the compressibility properties of strings
under any compression method, even if each string is compressed using a different compression. In this paper we set out to
answer this question by using \emph{straight-line programs}. These
provide a generic platform for representing many popular compression schemes
including the LZ-family, Run-Length Encoding, Byte-Pair Encoding, and dictionary methods.

\hspace{10pt} For two strings of total length $N$ having
straight-line program representations of total size $n$, we
present an algorithm running in $O(n^{1.4}N^{1.2})$ time for
computing the edit-distance of these two strings under any
rational scoring function, and an $O(n^{1.34}N^{1.34})$-time
algorithm for arbitrary scoring functions. This improves on a
recent algorithm of Tiskin that runs in $O(nN^{1.5})$ time, and
works only for rational scoring functions. 

\vspace*{-5mm}
\end{abstract}
\maketitle

\section{Introduction}
\label{Section: Introduction}

The \emph{edit distance} between two strings over a fixed
alphabet $\Sigma$ is the minimum cost of transforming one
string into the other via a sequence of character deletion,
insertion, and replacement operations~\cite{WagnerFischer1974}.
The cost of these elementary editing operations is given by
some scoring function which induces a metric on strings over
$\Sigma$. The simplest and most common scoring function is the
Levenshtein distance~\cite{Levenshtein1966} which assigns a
uniform score of 1 for every operation. Determining the
edit-distance between a pair of strings is a fundamental
problem in computer science in general, and in combinatorial
pattern matching in particular, with applications ranging from
database indexing and word processing, to
bioinformatics~\cite{Gusfield1997}.

The standard dynamic programming solution for computing the
edit distance between a pair of strings $A=a_1 a_2 \cdots a_N$
and $B=b_1 b_2 \cdots b_N$ involves filling in an $(N+1) \times
(N+1)$ table $T$, with $T[i,j]$ storing the edit distance
between $a_1 a_2 \cdots a_i$ and $b_1 b_2 \cdots b_j$. The
computation is done according to the base-case rules given by
$T[0,0] = 0$, $T[i,0] = T[i-1,0] + $ the cost of deleting
$a_i$, and $T[0,j] = T[0,j-1] + $ the cost of inserting $b_j$,
and according to the following dynamic~programming~step:
\begin{equation}
\label{Equation: Edit distance DP}%
T[i,j] = \min
\begin{cases}
T[i-1,j] \text{ + the cost of
deleting $a_i$}\\
T[i,j-1] \text{ + the cost of
inserting $b_j$}\\
T[i-1,j-1] \text{ + the cost of
replacing $a_i$ with $b_j$}\\
\end{cases}
\end{equation}
Note that as $T$ has $(N+1)^2$ entries, the time-complexity of
the algorithm above is $O(N^2)$. 

Compression is traditionally used to efficiently store data. In this paper, we focus
on using compression to accelerate the dynamic-programming solution for the
edit-distance problem described above. The basic idea is to first compress the
strings, and then compute the edit distance between the compressed strings. Note
that the ``acceleration via compression" approach has been successfully applied
also to other classical problems on strings. Various compression schemes, such as
LZ77~\cite{ZivLempel1977}, LZW-LZ78~\cite{ZivLempel1976}, Huffman coding, Byte-Pair
Encoding (BPE)~\cite{Shibata-et-al-1999}, Run-Length Encoding (RLE), were employed to accelerate exact string matching~\cite{AmirLandauSokol2003,KarkkainenUkkonen1996,Lifshits2007,Manber1994,Shibata-et-al-2000}, subsequence matching~\cite{CegielskiGuessarianLifshitsMatiyasevich2006},
approximate pattern matching~\cite{AmirBensonFarach1996,KarkkainenNavarroUkkonen2000,KarkkainenUkkonen1996,NavarroKidaetal2001}, and more
~\cite{MozesWeimannZiv2007}.

Regarding edit-distance computation, Bunke and Csirik presented a simple algorithm
for computing the edit-distance of strings that compress well under
RLE~\cite{BunkeCsirik1995}. This algorithm was later improved in a sequence of
papers~\cite{ApostolicoLandauSkiena1999,ArbellLandauMitchell2001,CrochemoreLandauZiv-Ukelson2003,MakinenNavarroUkkonen1999}
to an algorithm running in time $O(nN)$, for strings of total length $N$ that encode
into run-length strings of total length $n$.
In~\cite{CrochemoreLandauZiv-Ukelson2003}, an algorithm with the same time
complexity was given for strings that are compressed under LZW-LZ78, where $n$ again
is the length of the compressed strings. Note that this algorithm is also $O(N^2 /
\lg N)$ in the worst-case for any strings over constant-size alphabets.

The first paper to break the quadratic time-barrier of
edit-distance computation was the seminal paper of Masek and
Paterson~\cite{MasekPaterson1980}, who applied the
"Four-Russians technique" to obtain a running-time of
$O({N^2}/{\lg N})$ for any pair of strings, and of $O({N^2}/{\lg^2 N})$ assuming a unit-cost RAM model. Their algorithm
essentially exploits repetitions in the strings to obtain the
speed-up, and so in many ways it can also be viewed as
compression-based. In fact, one can say that their algorithm
works on the ``naive compression" that all strings over
constant-sized alphabets have. A drawback of the the Masek and
Paterson algorithm is that it can only be applied when the
given scoring function is rational. That is, when all costs of
editing operations are rational numbers. Note that this
restriction is indeed a limitation in biological applications,
where PAM and evolutionary distance similarity matrices are used for
scoring~\cite{CrochemoreLandauZiv-Ukelson2003,MasekPaterson1980}.
For this reason, the algorithm
in~\cite{CrochemoreLandauZiv-Ukelson2003} mentioned above was
designed specifically to work for arbitrary scoring functions.
We mentioned also Bille and Farach-Colton~\cite{BilleFarach2005} who extend
the Masek and Paterson algorithm to general alphabets.

There are two important things to observe from the above:
First, all known techniques for improving on the $O(N^2)$ time
bound of edit-distance computation, essentially apply
acceleration via compression. Second, apart from RLE, LZW-LZ78,
and the naive compression of the Four-Russians technique, we do
not know how to efficiently compute edit-distance under other
compression schemes. For example, no algorithm is known which
substantially improves $O(N^2)$ on strings which compress well
under LZ77. Such an algorithm would be interesting since there
are various types of strings that compress much better under
LZ77 than under RLE or LZW-LZ78. In light of this, and due to
the practical and theoretical importance of substantially
improving on the quadratic lower bound of string edit-distance
computation, we set out to answer the following question:
\begin{quote}
``Is there a general compression-based edit-distance algorithm
that can exploit the compressibility of two strings under
\emph{any} compression scheme?''
\end{quote}
A key ingredient to answering this question, we believe, lies
in a notion borrowed from the world of formal languages: The
notion of straight-line programs.

\subsection{Straight-line programs}
\label{Subsection: Straight-line programs}

A \emph{straight-line program} (SLP) is a context-free grammar
generating exactly one string. Moreover, only two types of
productions are allowed: $X_i\rightarrow a$ where $a$ is a
unique terminal, and $X_i\rightarrow X_pX_q$ with \mbox{$i>
p,q$} where $X_1,\ldots, X_n$ are the grammar variables. Each
variable appears exactly once on the left hand side of a
production. The string represented by a given SLP is a unique
string corresponding to the last nonterminal $X_n$. We define
the size of an SLP to be $n$, the number of variables (or
productions) it has. The length of the strings that is
generated by the SLP is denoted by $N$. It is important to
observe that many SLPs can be exponentially smaller than the
string they generate.

\begin{example}
Consider the string $abaababaabaab$. It could be generated by
the following SLP, also known as the \emph{Fibonacci SLP}:\\\\
\medskip
\qquad \qquad \parbox{5cm}{ $X_1\rightarrow b$
 \par $X_2\rightarrow a$
 \par $X_3\rightarrow X_2X_1$
 \par $X_4\rightarrow X_3X_2$
 \par $X_5\rightarrow X_4X_3$
 \par $X_6\rightarrow X_5X_4$
 \par $X_7\rightarrow X_6X_5$
}
 \parbox{8cm}{
\includegraphics[scale=.5]{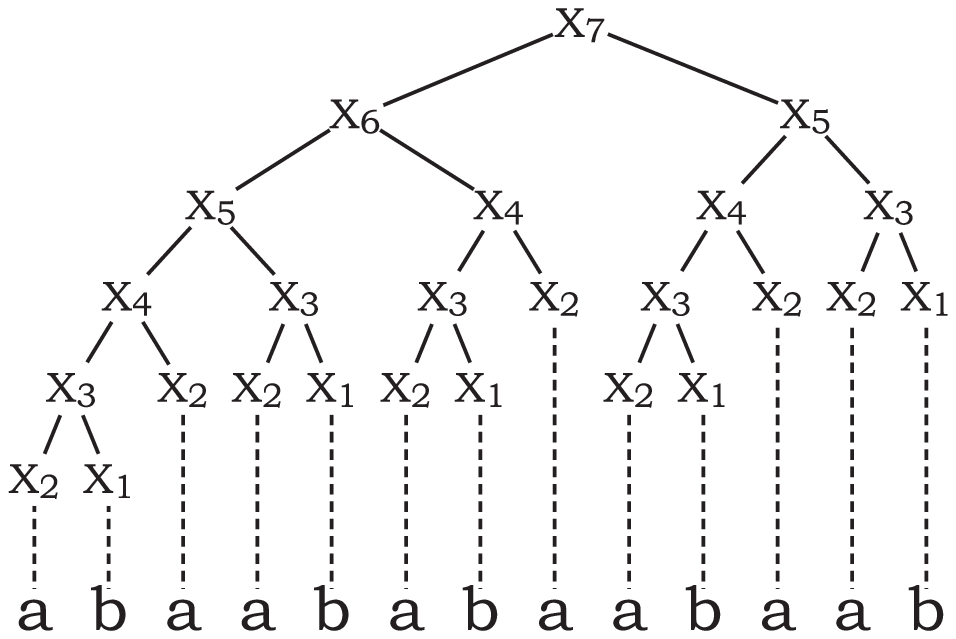}
}
\end{example}

Rytter~\cite{Rytter2003} proved that the resulting encoding of
most compression schemes including the LZ-family, RLE, Byte-Pair Encoding, and
dictionary methods, can be transformed to straight-line
programs quickly and without large expansion\footnote{Important exceptions of this list are statistical compressors such as Huffman or arithmetic coding, as well as compressions that are applied after a Burrows-Wheeler transformation.}. In particular, consider an LZ77
encoding~\cite{ZivLempel1977} with $n'$ blocks for a string of length
$N$. Rytter's algorithm produces an SLP-representation with
size $n = O(n' \log N)$ of the same string, in $O(n)$ time.
Moreover, $n$ lies within a $\log N$ factor from the size of a
\emph{minimal} SLP describing the same string. This gives us an
efficient logarithmic approximation of minimal SLPs, since
computing the LZ77 encoding of a string can be done in
linear-time. Note also that any string compressed by the
LZ78-LZW encoding can be transformed directly into a
straight-line program within a constant factor.

\subsection{Our results}
\label{Subsection: Our results}

Due to Rytter's results, SLPs are perfect candidates for
achieving our goal of generalizing compression-based
edit-distance algorithms. Indeed, a fast edit-distance
algorithm for strings that have small SLP representations,
would give a fast algorithm for strings which compress well
under the compression schemes generalized by SLPs. Note that
since constructing the strings generated by the SLPs requires
linear-time in the length of the strings, an $O(N^2)$ algorithm
is available via the standard dynamic-programming formulation
(\ref{Equation: Edit distance DP}). The main result of this
paper gives an algorithm which beats~this~bound:
\begin{theorem}
\label{Theorem: Main Result}%
Let $\mathcal{A}$ and $\mathcal{B}$ be two SLPs of total size
$n$ that respectively generate two string $A$ and $B$ of total
size $N$. Then, given $\mathcal{A}$ and $\mathcal{B}$, one can
compute the edit-distance between $A$ and $B$ in
$O(n^{1.4}N^{1.2})$ time for any rational scoring function.
\end{theorem}
We can remove the dependency of rational scoring schemes in
Theorem~\ref{Theorem: Main Result}, recalling that arbitrary
scoring schemes are important for biological applications. We
obtain the following secondary result for arbitrary scoring
functions:
\begin{theorem}
\label{Theorem: Secondary Result}%
Let $\mathcal{A}$ and $\mathcal{B}$ be two SLPs of total size
$n$ that respectively generate two string $A$ and $B$ of total
size $N$. Then, given $\mathcal{A}$ and $\mathcal{B}$, one can
compute the edit-distance between $A$ and $B$ in
$O(n^{1.34}N^{1.34})$ time for any arbitrary scoring function.
\end{theorem}


In the last part of the paper, we explain how the four-russians
technique can also be incorporated into our SLP edit-distance
scheme. We obtain a very simple algorithm that matches the
performance of~\cite{CrochemoreLandauZiv-Ukelson2003} in the
worst-case. That is, we obtain a four-russian like algorithm
with an $\Omega(\lg N)$ speed-up which can handle arbitrary
scoring functions, unlike the Masek and Paterson algorithm
which works only for rational functions. We add this algorithm
to our presentation not only for its practical importance, but
also to emphasize the fact that SLPs provide a framework which
allows an almost perfect generalization of compression-based
edit-distance algorithms.

\subsection{Related Work}
\label{Subsection: Related Work}

Rytter \emph{et al.}~\cite{KarpinskiRytterShinohara1995} was the first to consider
SLPs in the context of pattern matching, and other subsequent papers also followed
this line~\cite{LehmanShelat2002,MiyazakiShinoharaTakeda1997}. In~\cite{Rytter2003}
and~\cite{Lifshits2007} Rytter and Lifshits took this work one step further by
proposing SLPs as a general framework for dealing with pattern matching algorithms
that are accelerated via compression. However, the focus of Lifshits was on
determining whether or not these problems are polynomial in $n$ or not. In
particular, he gave an $O(n^3)$-time algorithm to determine equality of
SLPs~\cite{Lifshits2007}, and he established hardness for the edit
distance~\cite{LifshitsLohrey2006}, and even for the hamming distance
problems~\cite{Lifshits2007}. Nevertheless, Lifshits posed as an open problem the
question of whether or not there is an $O(nN)$ edit-distance algorithm for SLPs.
Here, our focus is on algorithms which break the quadratic $O(N^2)$ time-barrier,
and therefore all algorithms with running-times between $O(nN)$ and $O(N^2)$ are
interesting for us.

Recently, Tiskin~\cite{Tiskin2008} gave an $O(nN^{1.5})$ algorithm for computing the
longest common subsequence between two SLPs, an algorithm which can be extended at
constant-factor cost to compute the edit-distance between the SLPs under any
rational scoring function. Observe that our algorithm for arbitrary scoring
functions in Theorem~\ref{Theorem: Secondary Result} is already faster than Tiskin's
algorithm for most values of $N$ and $n$. Also, it has the advantage of being much
more simpler to implement. As for our main algorithm of Theorem~\ref{Theorem: Main
Result}, our faster running-time is achieved also by utilizing some of the
techniques used by Tiskin in a more elaborate way.

\section{The $\boldmath{DIST}$ Table}
\label{Section: The DIST Table}

The central dynamic-programming tool we use in our algorithms is the $DIST$ table, a
simple and handy data-structure which was originally introduced by Apostolico
\emph{et al.}~\cite{ApostolicoAtallahLarmoreMcFaddin1990}, and then further
developed by others in~\cite{CrochemoreLandauZiv-Ukelson2003,Schmidt1998}. In the
following section we briefly review basic facts about this tool that are essential
for understanding our results, following mostly the presentation
in~\cite{CrochemoreLandauZiv-Ukelson2003}.  We begin with the so-called
dynamic-programming grid, a graph representation of edit-distance computation on
which $DIST$ tables are defined.

\begin{figure}[h!]
\begin{center}
\includegraphics[scale=0.8]{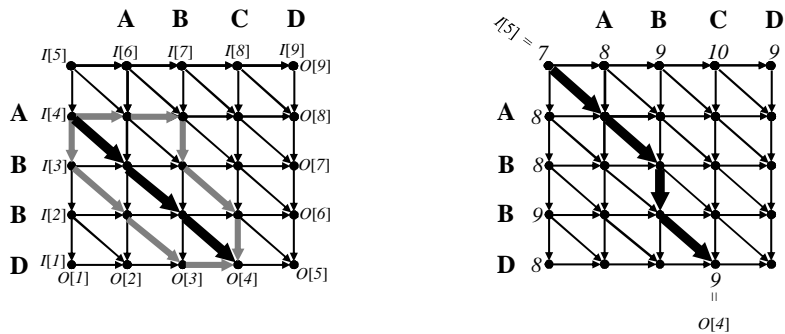}
\end{center}
\caption{A subgraph of a Levenshtein distance dynamic program graph. On the left, $DIST[4,4]$ (in bold) gives the minimum-weight path from $I[4]$ to $O[4]$. On the right, the value 9 of $O[4]$ is computed by $min_{i} I[i] + DIST[i,4]$.} \label{Figure:DIST}
\end{figure}

Consider the standard dynamic programming formulation
(\ref{Equation: Edit distance DP}) for computing the  edit-distance between two strings $A=a_1 a_2 \cdots a_N$
and $B=b_1 b_2 \cdots b_N$. The \emph{dynamic-programming grid} associated with this program, is an acyclic-directed graph which has a vertex for each entry of $T$ (see Figure~\ref{Figure:DIST}). The vertex corresponding to $T[i,j]$ is associated with $a_i$ and $b_j$, and has incoming edges according to (\ref{Equation: Edit distance DP}) -- an edge from $T[i-1,j]$ whose weight is the cost of deleting $a_i$, an edge from $T[i,j-1]$ whose weight is the cost of inserting $b_j$,
and an edge from $T[i-1,j-1]$ whose weight is the cost of replacing $a_i$ with $b_j$. The \emph{value} at the vertex corresponding to $T[i,j]$ is the value stored in $T[i,j]$, \emph{i.e.} the edit-distance between the length $i$ prefix of $A$ and the length $j$ prefix of $B$. Using the dynamic-programming grid $G$, we reduce the problem of computing the edit-distance between $A$ and $B$ to the problem of computing the weight of the lightest path from the upper-left corner to bottom-right corner in $G$.

We will work with sub-grids of the dynamic-programming grid that will be referred to as \emph{blocks}. The \emph{input vertices} of a block are all vertices in the first row and column of the block, while its \emph{output vertices} are all vertices in the last row and column. Together, the input and output vertices are referred to as the \emph{boundary} of the block. The substrings of $A$ and $B$ associated with the block are defined in the straightforward manner according to its first row and column. Also, for convenience purposes, we will order the input and output vertices, with both orderings starting from the vertex in bottom-leftmost corner of the block, and ending at the vertex in the upper-rightmost corner. The $i$th input vertex and $j$th output vertex are the $i$th and $j$th vertices in these orderings. We next give the definition of $DIST$ tables, defined over blocks of $G$.

\begin{definition}[$\boldmath{DIST}$~\cite{ApostolicoAtallahLarmoreMcFaddin1990}]
Let $G'$ be a block in $G$ with $x$ input vertices and $x$ output vertices.  The \emph{\textrm{DIST} table} corresponding to $G'$ is an $x \times x$ matrix, with $DIST[i,j]$ storing the weight of the minimum-weight path from the $i$th input to the $j$th output in $G$, and otherwise $\infty$ if no such paths exists.
\end{definition}

It is important to notice that the values at the output vertices of a block are completely determined by that values at its input and its corresponding $DIST$ table. In particular, if $I[i]$ and $O[j]$ are the values at the $i$th input vertex and $j$th output vertex of a block $G'$ of $G$, then
\begin{equation}
\label{Equation: Block Output}%
O[j] = \min_{1 \leq i \leq x} I[i] + DIST[i,j].
\end{equation}
Equation~\ref{Equation: Block Output} implies not only the
input-output relation of the dynamic-programming values of a
block, but also that the values at the output vertices can be
computed in linear time from the values at the input vertices.
Indeed, by (\ref{Equation: Block Output}), the values at the output vertices of $G'$ are given by the column minima of the matrix $I + DIST$. Furthermore, by a simple modification of
all $\infty$ values in $I + DIST$, we get what is known as a
\emph{totally-monotone matrix}~\cite{CrochemoreLandauZiv-Ukelson2003}. Now, Aggarwal
\emph{et al.}~\cite{AggarwalKlaweMoranShorWilber1987} gave a simple recursive algorithm, nicknamed SMAWK in the literature, that computes all column minima of an $x \times x$ totally-monotone matrix by querying only $O(x)$ elements of the matrix. It follows that using SMAWK we can compute the output values of $G'$ in $O(x)$ time.


Let us now discuss how to efficiently construct the $DIST$ table corresponding to a block in $G$. Observe that this can be done quite easily in $O(x^3)$ time, for blocks with boundary size $O(x)$, by computing the standard dynamic-programming table between every prefix of $A$ against $B$ and every prefix of $B$ against $A$. Each of these dynamic-programming tables contains all values of a particular row in the $DIST$ table. In~\cite{ApostolicoAtallahLarmoreMcFaddin1990}, Apostolico \emph{et al.} show an elegant way to reduce the time-complexity of this construction to $O(x^2 \lg x)$. In the case of rational scoring functions, the complexity can be further reduced to $O(x^2)$ as shown by Schmidt~\cite{Schmidt1998}.


\section{Acceleration via Straight-Line Programs}
\label{Section: Acceleration via Straight-Line Programs}

In the following section we describe a generic framework for
accelerating the edit distance computation of two strings which
are given by their SLP representation. This framework will
later be used for explaining all our algorithms. We will refer
throughout the paper to this framework as the \emph{block
edit-distance} procedure.

Let $\mathcal{A}$ and $\mathcal{B}$ be two SLP representations of a pair of strings
$A$ and $B$, and for ease of presentation assume that
$|\mathcal{A}|=|\mathcal{B}|=n$ and $|A|=|B|=N$. Recall the definition in
Section~\ref{Section: The DIST Table} for the dynamic-programming grid corresponding
to $A$ and $B$. The general idea behind the block edit-distance procedure is to
partition this grid into disjoint blocks, and then to compute the edit-distance
between $A$ and $B$ at the cost of computing the values at the boundary vertices of
each block. This is achieved by building in advance a repository containing all
$DIST$ tables corresponding to blocks in the partition. To efficiently construct this
repository, we show how to partition the grid in a way which induces many block
repeats. This is possible by utilizing substring repeats in $A$ and $B$ that are
captured in $\mathcal{A}$ and $\mathcal{B}$, and imply block repeats in the
partitioning of $G$. The edit-distance of $A$ and $B$ is then computed by
propagating the dynamic programming values at the boundary vertices of the blocks
using the $DIST$ tables in the repository and SMAWK. Before giving a complete
description of this algorithm, we need to introduce the notion of $xy$-partition.


\begin{definition}[$xy$-partition]
An $xy$-partition is a partitioning of $G$ into disjoint blocks
such that every block has boundary of size $O(x)$, and there
are $O(y)$ blocks in each row and column. In addition, we
require each pair of substrings of $A$ and $B$ associated with
a block to be generated by a pair of SLP variables in
$\mathcal{A}$ and $\mathcal{B}$.
\end{definition}

\begin{figure}[h!]
\begin{center}
\parbox{5cm}{\includegraphics[scale=0.45]{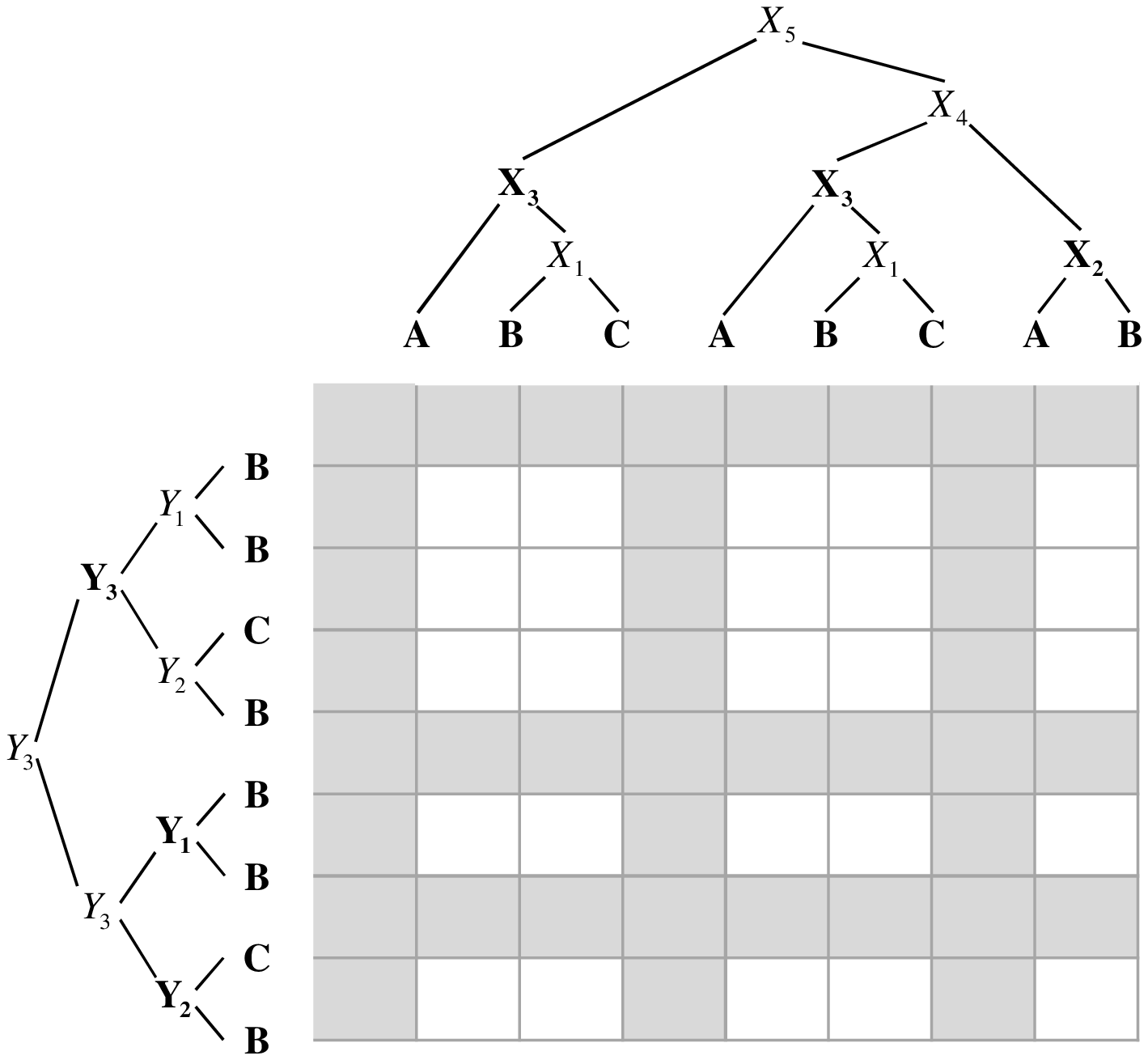}}
\parbox{2cm}{\ \ \   }
\parbox{7cm}{ \vspace{2cm} An $xy$-partition of an edit distance graph for two SLPs generating the strings ``ABCABCAB'' and ``BBCBBBCB''. The white blocks are the ones of the partition and their corresponding SLP variables are marked in bold. Notice that there are nine blocks in the partition but only six of them are distinct.}
\end{center}
\caption{An $xy$-partition.}
\label{Figure: x-block construction}
\end{figure}

An $xy$-partition of $G$ is a partition with a specific structure, but more importantly, one where each substring is generated by a unique SLP variable of $\mathcal{A}$ and $\mathcal{B}$. This latter requirement allows us to exploit the repetitions of $A$ and $B$ captured by their SLPs. We next give a complete description of the block edit distance procedure. It assumes an $xy$-partition of $G$ has already been constructed. Section~\ref{Section:
Constructing an xy-partition} explains how to construct such partitions.\\

\noindent \textbf{Block Edit Distance}
\begin{enumerate}
\item Construct a repository with the $DIST$ tables
    corresponding to each block in the $xy$-partition.
\item Fill-in the first row and column of $G$ using the
    standard base-case rules.
\item In top-to-bottom and left-to-right manner, identify
    the next block in the partition of $G$ and use its
    input and the repository to compute its output using
    (\ref{Equation: Block Output}).
\item Use the outputs in order to compute the inputs of the
    next blocks using (\ref{Equation: Edit distance DP}).
\item The value in the bottom-rightmost cell is the edit
    distance of $A$ and $B$.
\end{enumerate}

Apart from the repository construction in step 1, all details necessary for
implementing the block edit-distance procedure are by now clear. Indeed, steps 2 and
5 are trivial, and step 4 is done via the standard dynamic-programming formulation
of (\ref{Equation: Edit distance DP}). Furthermore, the SMAWK computation of output
values of a block, given its input values plus its corresponding $DIST$ table (step
3), is explained in Section~\ref{Section: The DIST Table}. We next show that, as we
are working with $xy$-partitions where each block is associated with an SLP
variable, we can compute a repository containing all $DIST$ necessary which is
rather small.

The first crucial observation for this, is that any two blocks
associated with the same pair of substrings $A'$ and $B'$ have
the same $DIST$ table. This is immediate since any such pair of
blocks have identical edge-weights.
\begin{observation}
A pair of substrings $A',B'$ uniquely identify the $DIST$ table
of a block.
\end{observation}
Since we required each substring in the $xy$-partition of $G$
to be generated by some SLP variable, the above observation
actually suggests that the number of different $DIST$ tables is
bounded by the number of variable pairs $X\in \mathcal{A}$ and
$Y\in \mathcal{B}$:
\begin{observation}
The number of different $DIST$ tables corresponding to any
$xy$-partition is $O(n^2)$.
\end{observation}
Therefore, combining the two observations above, we know that a
repository containing a $DIST$ tables for each SLP variable
pair $X\in \mathcal{A}$ and $Y\in \mathcal{B}$ will not be too
large, and that it will contain a table corresponding to each
block in our given $xy$-partition at hand. We can therefore
state the following lemma:

\begin{lemma}
\label{Lemma: xy arbitrary}%
The block edit-distance procedure runs in $O(n^2x^2\lg x + Ny)$
time.
\end{lemma}

\begin{proof}
We analyze the time complexity of each step in the block edit-distance procedure
separately. Step 1 can be performed in $O(n^2x^2\lg x)$ time, as we can construct
every $DIST$ table in $O(x^2\lg x)$ time (see Section~\ref{Section: The DIST
Table}), and the total number of such distinct matrices is $O(n^2)$. Step 2 can be done
trivially in $O(N)$ time. Then, step 3 takes $O(x)$ time per block by using the
SMAWK algorithm as explained in Section~\ref{Section: The DIST Table}. Step 4 also
takes $O(x)$ time per block as it only computes the values in the $O(x)$ vertices
adjacent to the output vertices. The total time complexity of steps 3 and 4 is thus
equal to the total number of boundary vertices in the $xy$-partition of $G$, and
therefore to $O(Ny)$. Accounting for all steps together, this gives us the time
complexity stated in the lemma.
\end{proof}

\section{Constructing an $\boldmath{xy}$-partition}
\label{Section: Constructing an xy-partition}

In this section we discuss the missing component of
Section~\ref{Section: Acceleration via Straight-Line Programs},
namely the construction of $xy$-partitions. In particular, we
complete the proof of Theorem~\ref{Theorem: Secondary Result}
by showing how to efficiently construct an $xy$-partition where
$y=O(nN/x)$ for every $x\le N$. Together with Lemma~\ref{Lemma:
xy arbitrary}, this implies an $O(n^{\frac{4}{3}}
N^{\frac{4}{3}}\lg^{\frac{1}{3}} N ) = O(n^{1.34}N^{1.34})$
time algorithm for arbitrary scoring functions by considering
$x=N^{\frac{2}{3}}/(n\lg N)^{\frac{1}{3}}$. In the remainder of
this section we prove the following lemma.

\begin{lemma}
\label{bomba shel lemma}%
For every $x\le N$ there exists an $xy$-partition with
$y=O(nN/x)$. Moreover, this partition can be found in $O(N)$
time.
\end{lemma}

To prove the lemma, we show that for every SLP $\mathcal{A}$
generating a string $A$ and every $x\le N$, one can partition
$A$ into $O(nN/x)$ disjoint substrings, each of length $O(x)$,
such that every substring is generated by some variable in
$\mathcal{A}$. This defines a subset of variables in both input
SLPs which together defined our desired $xy$-partition. To
partition $A$, we first identify $O(N/x)$ grammar variables in
$\mathcal{A}$ each generating a disjoint substring of length
between $x$ and $2x$. We use these variables to partition $A$.
We then show that the substrings of $A$ that are still not
associated with a variable can each be generated by $O(n)$
additional variables. Furthermore, these $O(n)$ variables each
generate a string of length bounded by $x$. We add all such
variables to our partition of $A$ for a total of $O(nN/x)$
variables.

Consider the parse tree of $\mathcal{A}$. We want to identify
$O(nN/x)$ \emph{key-vertices} such that every key-vertex
generates a substring of length $O(x)$, and $A$ is a
concatenation of substrings generated by key-vertices. We start
by marking every vertex $v$ that generates a substring of
length greater than $x$ as a key-vertex iff both children of
$v$ generate substrings of length smaller than $x$. This gives
us $\ell \leq N/x$ key-vertices so far, each generating a
substring of length $\Theta(x)$ (see~Figure~\ref{Figure:
keyvertices}). But we are still not guaranteed that these
vertices cover $A$ entirely.

\begin{figure}[t]
\begin{center}
\parbox{5cm}{\includegraphics[scale=0.55]{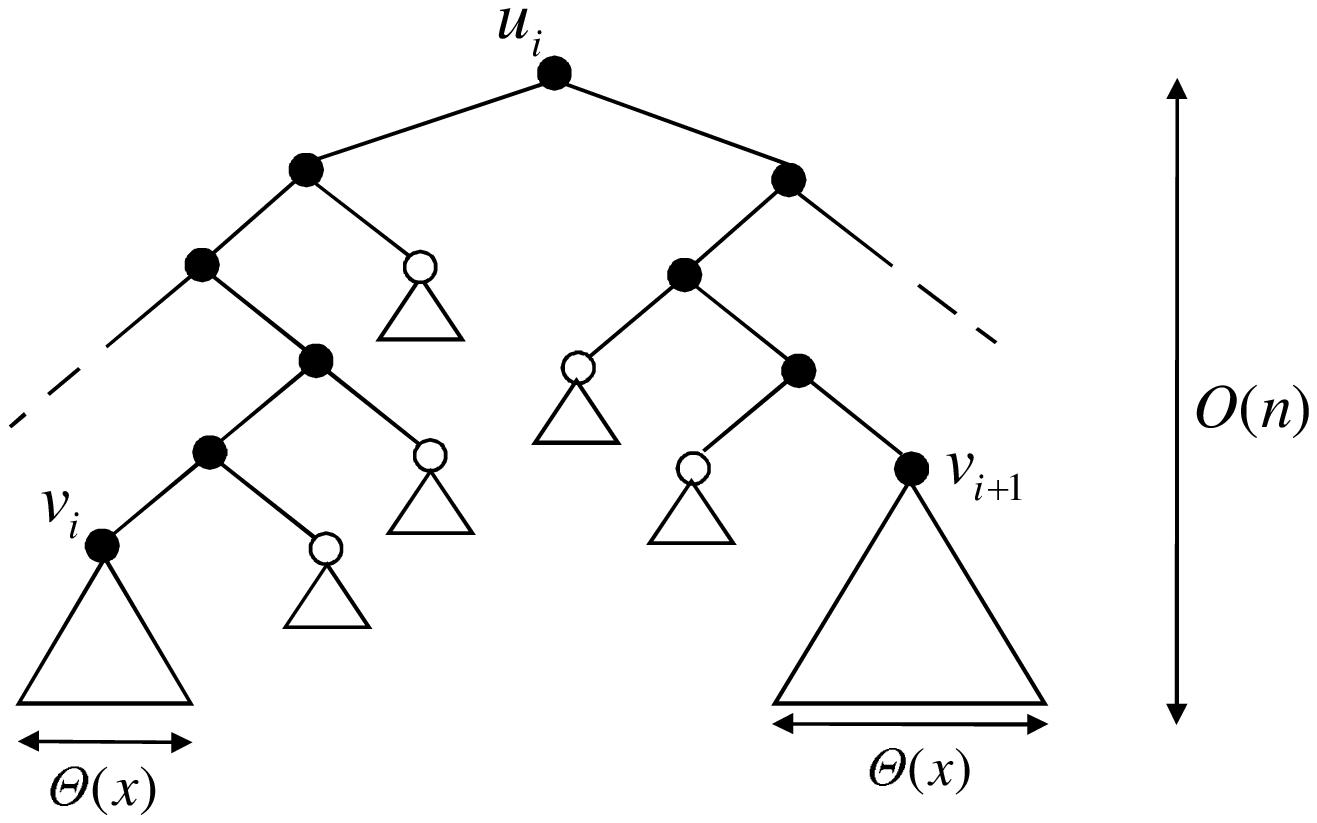}}
\parbox{3cm}{\ \ \   }
\parbox{6cm}{ The key-vertices
$v_i$ and $v_{i+1}$ both generate substrings of length $\Theta(x)$, and their least common ancestor is $u_i$. The white vertices ``hanging'' of the $v_i$-to-$v_{i+1}$ path are the added key-vertices. Together, they generate the substring that lies between the substrings of $v_i$ and $v_{i+1}$.}
\end{center}
\caption{A closer look on the parse tree of an SLP $\mathcal{A}$.}
 \label{Figure: keyvertices}
\end{figure}

To fix this, consider the ordering $v_1,v_2,\ldots,v_{\ell}$ on
the current key-vertices induced by a left-to-right postorder
traversal of the parse tree. This way, $v_{i+1}$ is ``to the
right of'' $v_i$. If every $v_i$ generates the substring $A_i$
then $A= A'_1A_1A'_2A_2\cdots A'_{\ell}A_\ell A'_{\ell+1}$,
where every $A_i$ is of length $\Theta(x)$, and every $A'_i$ is the ``missing'' substring of $A$ that lies between $A_{i-1}$ and $A_i$. We now show that
every $A'_i$ is a concatenation of substrings of length smaller than $x$
generated by at most $O(n)$ vertices.

Let $u_i$ be the lowest common ancestor of $v_i$ and $v_{i+1}$
and let $P_i$ (resp. $P_{i+1}$) be the unique path between
$u_i$ and $v_i$ (resp. $v_{i+1}$). For every vertex $v\in P_i
-\{u_i\}$ such that $v$'s left child is also in $P_i$ mark
$v$'s right child as a key-vertex. Similarly, for every vertex
$ v\in P_{i+1}-\{u_i\}$ such that $v$'s right child is also in
$P_{i+1}$ mark $v$'s left child as a key-vertex. It is easy to
verify that $A'_i$ is the concatenation of substrings generated
by these newly marked key-vertices. There are at most $2n$ of
these key-vertices since the depth of the parse tree is bounded
by the number of different SLP variables. Moreover, they each
generate a substring of length smaller than $x$ for the following reason. Assume for  contradiction that one of them generates a string of length greater than $x$. This would imply the existence of some vertex between $v_i$ and $v_{i+1}$ in the $v_1,v_2,\ldots,v_{\ell}$ ordering.

To conclude, we showed that $A= A'_1A_1A'_2A_2\cdots
A'_{\ell}A_\ell A'_{\ell+1}$ where $\ell \leq N/x$, every $A_i$ is of length
$\Theta(x)$ and is generated by one vertex, and every $A'_i$ is
a concatenation of $O(n)$ substrings each of length smaller than $x$
and generated by one vertex. Overall, we get that
$y=O(n\ell)=O(nN/x)$ vertices suffice to generate $A$ for every $x\le N$. It is easy to see that we can identify these vertices in $O(N)$ time thus proving Lemma~\ref{bomba shel lemma}. By choosing $x=N^{\frac{2}{3}}/(n\lg N)^{\frac{1}{3}}$, and using the block edit distance time complexity of Lemma~\ref{Lemma:
xy arbitrary}, this implies an $O(n^{1.34}N^{1.34})$
time algorithm for arbitrary scoring functions.

\section{Improvement for Rational Scoring Functions}
\label{Section: Improvement for Rational Scoring Functions}

In this section we show that in the case of rational scoring
functions, the time complexity of the block edit distance
procedure can be reduced substantially by using a recursive
construction of the $DIST$ tables. In particular, we complete
the proof of Theorem~\ref{Theorem: Main Result} by showing that
in this case the repository of $DIST$ tables can be computed in
$O(n^2x^{1.5})$ time. This implies an $O(n^{1.4}N^{1.2})$ time
algorithm for rational scoring functions by considering $x =
N^{0.8}/n^{0.4}$ and the $xy$-partition with $y=nN/x$.

Before we describe how to compute the repository in $O(n^2x^{1.5})$ time, we need
to introduce some features that $DIST$ tables over rational scoring functions have.
The first property, discovered by Schmidt~\cite{Schmidt1998}, is what is known as
the succinct representation property: Any $x\times x$ $DIST$ table can be succinctly
stored using only $O(x)$ space. This follows from considering the vector obtained by
subtracting a $DIST$ column from the column to its right, and observing that this
vector has only a constant number of value changes. The second property is that
succinct representations allow to efficiently merge two $DIST$ tables. That is, if
$D_1$ and $D_2$ are two $DIST$ tables, one between a pair of substrings $A'$ and
$B'$ and the other between $A'$ and $B''$, then we refer to the $DIST$ table between
$A'$ and $B'B''$ as the product of \emph{merging} $D_1$ and $D_2$. A recent
important result of Tiskin~\cite{Tiskin2006} shows how to utilize the succinct
representation of $DIST$ tables in order to merge two succinct $x\times x$ $DIST$
tables in $O(x^{1.5})$ time.

\begin{lemma}
\label{Lemma: xy rational}%
The block edit distance algorithm runs in $O(n^2x^{1.5} + Ny)$
time in case the underlying scoring function is rational.
\end{lemma}

\begin{proof}
To prove the lemma it suffices to show how to compute the
repository of $DIST$ tables in step 1 of the block
edit-distance procedure in $O(n^2x^{1.5})$ time, in case the
underlying scoring function is rational.
We will work with succinct representations of the $DIST$ tables
as described above. Say $X\rightarrow X_pX_q$ and $Y\rightarrow
Y_sY_t$ are two rules in the SLPs $\mathcal{A}$ and
$\mathcal{B}$ respectively. To compute the $DIST$ table that
corresponds to the strings generated by $X$ and $Y$, we first
recursively compute the four $DIST$ tables that correspond to
the pairs $(X_p,Y_s)$, $(X_p,Y_t)$, $(X_q,Y_s)$, and
$(X_q,Y_t)$. We then merge these four tables to obtain the
$DIST$ table that corresponds to $(X,Y)$. To do so we use
Tiskin's procedure to merge $(X_p,Y_s)$ with $(X_p,Y_t)$ into
$(X_p,Y_sT_t)$, then merge $(X_q,Y_s)$ with $(X_q,Y_t)$ into
$(X_q,Y_sT_t)$, and finally we merge $(X_p,Y_sT_t)$ and
$(X_q,Y_sT_t)$ into $(X_pX_q,Y_sT_t)=(X,Y)$. This recursive
procedure computes each succinct $DIST$ table by three merge
operations, each taking $O(x^{1.5})$ time and $O(x)$ space.
Since the number of different $DIST$ tables is bounded by
$O(n^2)$, the $O(n^2x^{1.5})$ time for constructing the
repository follows.
\end{proof}

To conclude, we have shown an $O(n^2x^{1.5} + Ny)$ time algorithm for computing the edit distance. Using the $xy$-partition from Lemma~\ref{bomba shel lemma} with $x =
N^{0.8}/n^{0.4}$ and $y=nN/x$, we get a time complexity of $O(n^{1.4}N^{1.2})$ .

\section{Four-Russian Interpretation}
\label{Section: Four-Russian Interpretation}

In the previous sections we showed how SLPs can be used to speed up the edit distance computation of strings that compress well under some compression scheme.
In this section, we conclude the presentation of our SLP framework by presenting an $\Omega(\lg N)$ speed-up for strings that do not compress well under any compression scheme. To do so, we adopt the Four Russions approach of Masek and Paterson~\cite{MasekPaterson1980} that utilizes a naive property that every string over a fixed alphabet has. Namely, that short enough substrings must appear many times. However, while the Masek and Paterson algorithm can only handle rational scoring functions, the SLP version that we propose can handle arbitrary scoring functions.

Consider a string $A$ of length $N$ over an alphabet $\Sigma$. The parse tree of the naive SLP $\mathcal{A}$ is a complete binary tree with $N$ leaves\footnote{We assume without loss of generality that $N$ is a power of 2.}. This way, for every $x\le N$ we get that $A$ is the concatenation of $O(N/x)$ substrings each of length $\Theta(x)$ and each can be generated by some variable in $\mathcal{A}$. This partition of $A$ suggests an $xy$-partition in which $y=N/x$. At first glance, this might seem better than the partition guarantee of Lemma~\ref{bomba shel lemma} in which $y=nN/x$. However, notice that in the naive SLP we have $n\ge N$ so we can not afford to compute a repository of $O(n^2)$ $DIST$ tables.

To overcome this problem, we choose $x$ small enough so that $|\Sigma|^x$, the number of possible substrings of length $x$, is small. In particular, by taking $x=\frac{1}{2} \log_{|\Sigma|} N$ we get that the number of possible substrings of length $x$ is bounded by $|\Sigma|^x=\sqrt{N}$. This implies an $xy$-partition in which $x=\frac{1}{2} \log_{|\Sigma|} N$, $y=N/x$, and the number of distinct blocks $n'$ is $O(N)$. 
Using this partition, we get that the total construction time of the $DIST$ repository is $O(n' x^2\lg x)$. 
Similar to Lemma~\ref{Lemma: xy arbitrary}, we get that the total running time of the block edit distance algorithm is $O(n' x^2\lg x+Ny)$ which gives $O(N^2/\lg N)$.


%
%

\bibliographystyle{plain}
\bibliography{biblio}

\end{document}